%% file: tkomatsubara_arXiv.tex
\documentclass[12pt]{article}
\usepackage{epsfig,graphicx}
\usepackage{fpcp04}

\input fpcp04-macro
\begin{document}

\Title{Rare Kaon Decays - a review of results}
\bigskip

%%%%%%%%%%%%%%%%%%%%%%%%%%%%%%%%%%%%%
% Label to flag the first page of your contribution
% Replace Perret by your name starting with a capital letter
%
\label{TKomatsubaraStart}

%%%%%%%%%%%%%%%%%%%%%%%%%%%%%%%%%%%%%
% Your name
%
\author{ Takeshi K. Komatsubara\index{TKomatsubara} }

%%%%%%%%%%%%%%%%%%%%%%%%%%%%%%%%%%%%%
% Your address
%
\address{Institute of Particle and Nuclear Studies\\
High Energy Accelerator Research Organization (KEK)\\
Oho 1-1, Tsukuba, Ibaraki 305-0801, JAPAN\\
Email: takeshi.komatsubara@kek.jp\\
}

\makeauthor\abstracts{
 Recent results and future prospects of rare kaon-decay experiments 
 at KEK, BNL, CERN and FNAL are reviewed. 
 Topics include
 lepton flavor violation, 
 T-violating transverse muon polarization in 
 {\ensuremath{K^+_{\mu 3}}\xspace},
 exotic decays,
 {\ensuremath{\Kz \to \piz \ellell}\xspace},
 and 
 {\ensuremath{\kaon \to \pi \nunub}\xspace}.
}

\section{Introduction}
The branching ratio (\BR) for a decay mode $f$ of a particle is written as 
\begin{equation}
  \BR(f) \equiv \frac{\Gamma_f}{\Gamma_{all}} = \tau \times \frac{2\pi}{\hbar} 
       \int d(phase~space) \cdot |M_{f}|^2 
\label{eq:tkomatsubara-decay}
\end{equation}
in the S-matrix version of Fermi's Golden Rule No.2.
\KS, \Kp, and \KL decays with the branching ratio of $10^{-10}$ 
correspond to the partial widths of 
$0.73\times 10^{-15}$, $5.3\times 10^{-18}$, and 
$1.3\times 10^{-18}$eV, respectively. 
Kaon decays at $10^{-7}$ or less are categorized as 
``rare'' decays~\footnote{
   In this conference, 
   ``medium'' and ``well-done'' kaon decays 
   were reviewed by F. Bossi~\cite{tkomatsubara-Bossi}
   and S. Glazov~\cite{tkomatsubara-Glazov} and
   theoretical issues were discussed by X-G He~\cite{tkomatsubara-He}. 
},
and are the frontiers that no other heavy-flavor physics can reach at 
present. 
The smallest branching ratio yet measured in particle physics is 
$(8.7^{+5.7}_{-4.1})\times 10^{-12}$
for the decay {\ensuremath{\KL \to \epem}\xspace}~\cite{tkomatsubara-E871ee}, 
and the most stringent upper limit is $<4.7 \times 10^{-12}$
for {\ensuremath{\KL \to \mu^{\pm} e^{\mp}}\xspace}~\cite{tkomatsubara-E871mue}; 
both of these were achieved by the E871 experiment at BNL. 
 
 There should be reasons why a decay is particularly rare; 
except for the trivial case that the phase space is too small, 
the decay amplitude  $M_{f}$ must be small because
1) $M_{f}$ is absolutely zero due to the symmetry in Theory of Everything, 
2) the decay does exist at a tree level but the mass of the intermediate boson 
   is too heavy, or
3) there is no tree diagram but are loop diagrams with suppression mechanisms
  such as GIM. 
We search for violations of the Standard Model (SM) in the second case, 
and study the flavor parameters and CP violations in and beyond the SM 
in the third case.

 In this article, recent results and future prospects of rare kaon-decay experiments
(table~\ref{tab:tkomatsubara-explist})
are reviewed. 
For more information about this research field, 
the review in PDG-2004~\cite{tkomatsubara-PDG2004litt}
and latest talks in this summer~\cite{tkomatsubara-Redlinger,tkomatsubara-Mikulec,tkomatsubara-Patera}
are recommended.

 Reminder: the experimental upper limits in this article are 
at 90\% confidence level.

%%%%%%%%%%%%%%%%%%%%%%%%%%%%%%%%%%%%%%%%%%%%%%%%%%%%%%%%%%%%%%%%%%%%%%%%%
%%
%%   use this format to include a LaTeX table  into your paper
%%
\begin{table}[htbp]
\begin{center}
  \begin{tabular}{|l|lr|l|l|}
 \hline
 Lab & \multicolumn{2}{|l|}{Accelerator} & Experiment & Kaon decay\\
 \hline
 KEK & PS  		&(12 GeV)	&E246 $^{\surd}$		& \Kp\ at rest\\
     &                 	&		&E391a				& \KL\\
 KEK-JAERI& J-PARC PS   &(50 GeV)       &LoI's $^{\ast}$                & \KL, \Kp\ at rest\\                
 BNL & AGS 		&(25 GeV)	&E787 $^{\surd}$\ /\ E949	& \Kp\ at rest\\
     &          	&		&E865 $^{\surd}$  		& \Kp\ in flight\\
     &    		&		&KOPIO $^{\ast}$		& \KL\\
 CERN& SPS		&(400 GeV)	&NA48/1 $^{\surd}$ 		& \KS\\
     &                  &               &NA48/3 $^{\ast}$               & \Kp\ in flight\\
 FNAL& Tevatron		&(800 GeV)	&KTeV $^{\surd}$    		& \KL\\
     & Main Injector	&(120 GeV)	&CKM-P940 $^{\ast}$		& \Kp\ in flight\\
 \hline
  \end{tabular}
\caption{Rare kaon-decay experiments being reviewed in this article.
     The DA$\Phi$NE-KLOE experiment~\cite{tkomatsubara-Bossi}
     is not included.
    ``${\surd}$'' means  data taking of the experiment is completed; 
    ``${\ast}$'' means  construction of the experiment is not started.
}
\label{tab:tkomatsubara-explist}
\end{center}
\end{table}
%%%%%%%%%%%%%%%%%%%%%%%%%%%%%%%%%%%%%%%%%%%%%%%%%%%%%%%%%%%%%%%%%%%%%%%%%

\section{Explicit violations of the Standard Model}

Results of the searches for explicit SM violations
are summarized in table~\ref{tab:tkomatsubara-violation}.
``FC'' is the unified confidence intervals by 
Feldman and Cousins~\cite{tkomatsubara-FC}.

%%%%%%%%%%%%%%%%%%%%%%%%%%%%%%%%%%%%%%%%%%%%%%%%%%%%%%%%%%%%%%%%%%%%%%%%%
%%
%%   use this format to include a LaTeX table  into your paper
%%
\begin{table}[htbp]
\begin{center}
  \begin{tabular}{|l|cc|cl|ll|}
 \hline
 Mode & 
        $N_{obs}$ & $n_{bgd}$     &
          Result & 
            $Stat$  &
              Experiment      & Ref. \\
 \hline
 {\ensuremath{\Kp \to \pip \mup \en}\xspace} &
        8         & 8.2$\pm$1.9   &
          $<2.2\times 10^{-11}$   &
            likelihood  & 
              E865-'98       & 
                                                         \\
                            &
                   &               &
          $<1.2\times 10^{-11}$   &
                        &  
              E865 +E777  & 
                                  \cite{tkomatsubara-Sher}\\
 \hline
 {\ensuremath{\KL \to \piz \mu^\pm e^\mp}\xspace} &
                  &               &
          $<3.37\times 10^{-10}$  &
            Poisson & 
              KTeV &  
                                  \cite{tkomatsubara-Bellavance}\\
\hline
 {\ensuremath{P_T}\xspace} in {\ensuremath{K^+_{\mu 3}}\xspace} &
                  &               &
          $|P_T|<0.0050$  &
                      & 
              E246 & 
                                  \cite{tkomatsubara-E246}\\
 \hline
 {\ensuremath{\Kp \to \pim \mup \mup}\xspace} &
        5         & 5.3           &
          $<3.0\times 10^{-9}$   &
            FC &
              E865            & 
                                  \cite{tkomatsubara-E865ex}\\
 {\ensuremath{\Kp \to \pim \ep \ep}\xspace} &
        0         & negligible    &
          $<6.4\times 10^{-10}$   &
            FC               & 
              E865            & 
                                  \cite{tkomatsubara-E865ex}\\
 \hline
 {\ensuremath{\Kp \to \pip \g}\xspace} &
        0         & negligible    &
          $<3.6\times 10^{-7}$   &
            FC                  & 
              E787               & 
                                  \cite{tkomatsubara-E787pig}\\
 \hline
 {\ensuremath{\Kp \to \pip X^0}\xspace} &
        1         &               &
          $<7.3\times 10^{-11}$   &
            FC                  &
              E949-'02 +E787  & 
                                  \cite{tkomatsubara-E949}\\
 \hline
  \end{tabular}
\caption{Summary of the searches for explicit SM violations.
         This table includes the number of observed events ($N_{obs}$), 
         background estimation ($n_{bgd}$), and the statistical technique used to 
         obtain the result ($Stat$).
         {\ensuremath{\Kp \to \pip X^0}\xspace} is discussed in the next section.}
\label{tab:tkomatsubara-violation}
\end{center}
\end{table}
%%%%%%%%%%%%%%%%%%%%%%%%%%%%%%%%%%%%%%%%%%%%%%%%%%%%%%%%%%%%%%%%%%%%%%%%%

\subsection{Lepton flavor violation}

 Experimental search for lepton flavor (LF) violation in kaon decays
 ({\ensuremath{\KL \to \mu^{\pm} e^{\mp}}\xspace}, 
  {\ensuremath{\Kp \to \pip \mup \en}\xspace}, 
   {\ensuremath{\KL \to \piz \mu^\pm e^\mp}\xspace})
has a long history. 
The kaon system is well suited to the investigation of 
LF-violating new processes
involving both quarks and charged leptons~\footnote{
 Assuming an additive quantum number for quarks and leptons in the same generation
  (``one'' for down-quark and electron, ``two'' for strange-quark and muon, ...),
 the net number is conserved in the LF-violating kaon decays.}
due to the high sensitivity achieved by experiments on this system.
The mass of a hypothetical gauge boson for the tree-level effects 
should be in the scale of a few hundred 
TeV/$c^2$~\cite{tkomatsubara-LFVtheory1,tkomatsubara-LFVtheory2}.
A drawback is that
the LF-violating processes 
induced by Supersymmetric loop effects 
are not as promising as in $\mu^+\to e^+\gamma$ decay 
and $\mu^{-}N\to e^{-}N$ conversion, 
because ``Super-GIM'' suppression mechanism is 
expected in both quark and lepton sectors~\cite{tkomatsubara-SUSY}.
Theoretical motivations for these decays are discussed in 
\cite{tkomatsubara-LFVtheory3,tkomatsubara-LFVtheory4,
      tkomatsubara-LFVtheory5,tkomatsubara-LFVtheory6}.

 Three-body decays {\ensuremath{\kaon \to \pi \mu e}\xspace} have to be explored 
 in spite of the phase-space disadvantage, because they are sensitive to 
 vector and scaler interactions. 
 The best upper limit \BR({\ensuremath{\Kp \to \pip \mup \en}\xspace})
 $<1.2\times 10^{-11}$ was set by the E865 collaboration 
 with their 1995-1998 data sets and the result 
 of an earlier experiment E777~\cite{tkomatsubara-E777}.
 In the analysis of 1998 data set, 
 eight observed events were examined with a general Likelihood
 from the combination of probability density functions 
 for {\ensuremath{\Kp \to \pip \mup \en}\xspace}; 
 these events were not consistent with the signal~\footnote{
   Reminder: since the likelihood analysis technique had been adopted,
   the background level $8.2\pm 1.9$ in 1998
   did not mean the search was background-limited. },
 and  
 an upper limit of 2.4 events was determined for calculating the limit
of the branching ratio 
 $<2.2\times 10^{-11}$.
 An upper limit \BR({\ensuremath{\KL \to \piz \mu^\pm e^\mp}\xspace})
 $<3.37\times 10^{-10}$ was set as a preliminary result
from the KTeV collaboration with their 1997 and 1999 data sets. 

 \subsection{T-violating transverse muon polarization in 
             {\ensuremath{K^+_{\mu 3}}\xspace}}

 In the {\ensuremath{\Kp \to \piz \mup \nu}\xspace} decay
 ({\ensuremath{K^+_{\mu 3}}\xspace}, 
  \BR=$(3.27\pm0.06)$\%~\cite{tkomatsubara-PDG2004}), 
 the transverse muon polarization $P_T$
 (the perpendicular component of the muon spin vector relative to the decay plane
 determined by the momentum vectors of muon and pion 
 in the $K^+$ rest frame) is a T-odd quantity
 and is an observable of CP violation.
 Any spurious effect from final-state interactions is small
 ($<10^{-5}$),
 because no charged particle other than muon exists in the final state.
 $P_T$ is almost vanishing 
($\sim 10^{-7}$)~\cite{tkomatsubara-Sanda} in the SM,
 while new sources of CP violation may give rise to $P_T$
 as large as $10^{-3}$.
 $P_T$ in {\ensuremath{K^+_{\mu 3}}\xspace} has therefore been regarded as
 a sensitive probe of non-SM CP violation,    
 and is a good example of looking beyond the SM
 by measuring a decay property with high statistics.

 The E246 collaboration measured the charged track and photons
 from \Kp\ decays at rest
 with the superconducting toroidal spectrometer 
 (consisting of 12 identical spectrometers arranged in 
  rotational symmetry),
 which enabled the experiment 
 to control possible sources of systematic uncertainties
 in polarization measurement. 
 A new improved value 
 $P_T=(-0.17\pm 0.23${\footnotesize (stat)}$\pm 0.11${\footnotesize (syst)})$\times 10^{-2}$
 was obtained by the total data sets of E246 from 1996 to 2000, 
 giving an upper limit $|P_T|<0.0050$.

 \subsection{Exotic decays}

 E865 reported the upper limits 
 \BR({\ensuremath{\Kp \to \pim \mup \mup}\xspace})
 $<3.0\times 10^{-9}$
 and \BR({\ensuremath{\Kp \to \pim \ep \ep}\xspace})
 $<6.4\times 10^{-10}$.
 The former decay is a neutrino-less ``double muon'' decay 
 by changing total lepton number by two~\cite{tkomatsubara-LittShrock1},
 and provides a unique channel to search for effects of 
 Majorana neutrinos in the second 
 generation of quarks and leptons~\cite{tkomatsubara-Zuber,tkomatsubara-LittShrock2}.

  The decay {\ensuremath{\Kp \to \pip \g}\xspace}
 is a spin 0$\to$0 transition with a real photon
 and is thus forbidden by angular momentum conservation; 
 this decay is also forbidden on gauge invariance grounds. 
 An experimental signature of exotic physics, 
 such as non-commutative QED
 and/or non-commutative SM~\cite{tkomatsubara-exotic},
 could appear in this decay mode. 
 The E787 collaboration reported a new upper limit 
 \BR({\ensuremath{\Kp \to \pip \g}\xspace})
 $<3.6\times 10^{-7}$.

\section{Flavor Changing Neutral Current processes}

Results of the studies of Flavor Changing Neutral Current(FCNC) processes
are summarized in table~\ref{tab:tkomatsubara-FCNC}.
 ``J'' is the technique based on likelihood-ratio and 
  modified Frequentist's approach, used to set limits on the Higgs mass
  from measurements at LEP, by Junk~\cite{tkomatsubara-Junk}.
For the discussions in this section,
the Wolfenstein parametrization
of the Cabibbo-Kobayashi-Maskawa(CKM) matrix 
with $\lambda \equiv \sin \theta_C \simeq 0.22$, $A$, $\rho$, and $\eta$ 
is used.

%%%%%%%%%%%%%%%%%%%%%%%%%%%%%%%%%%%%%%%%%%%%%%%%%%%%%%%%%%%%%%%%%%%%%%%%%
%%
%%   use this format to include a LaTeX table  into your paper
%%
\begin{table}[htbp]
\begin{center}
  \begin{tabular}{|l|cc|cl|ll|}
 \hline
 Mode & 
        $N_{obs}$ & $n_{bgd}$     &
          Result & 
            $Stat$  &
              Experiment      & Ref. \\
 \hline
 {\ensuremath{\KS \to \piz \ep \en}\xspace} &
         7        & 0.15$^{+0.10}_{-0.04}$  &
          $5.8^{+2.8}_{-2.3}${\footnotesize (stat)}$\pm 0.8${\footnotesize (syst)} $\times 10^{-9}$ &
            FC  &
              NA48/1      & 
                                  \cite{tkomatsubara-NA48pi0ee}\\
 {\ensuremath{\KS \to \piz \mup \mun}\xspace} &
         6        & 0.22$^{+0.18}_{-0.11}$  &
          $2.9^{+1.5}_{-1.2}${\footnotesize (stat)}$\pm 0.2${\footnotesize (syst)} $\times 10^{-9}$ &
            FC  &
              NA48/1      & 
                                  \cite{tkomatsubara-NA48pi0mm}\\
 \hline
 {\ensuremath{\KL \to \piz \ep \en}\xspace} &
         2        & 1.06$\pm$0.41 &
          $<5.1\times 10^{-10}$  &
            FC &
              KTeV-'97      & 
                                  \cite{tkomatsubara-KTeVpi0ee97}\\
                            &
         1        & 0.99$\pm$0.35 &
          $<3.5\times 10^{-10}$  &
            FC &
              KTeV-'99     & 
                                                           \\ 
                            &
                  &               &
          $<2.8\times 10^{-10}$  &
                     & 
              \ \ combined &  
                                  \cite{tkomatsubara-KTeVpi0ee99}\\
 {\ensuremath{\KL \to \piz \mup \mun}\xspace} &
         2        & 0.87$\pm$0.15 &
          $<3.8\times 10^{-10}$  &
            FC    &
              KTeV-'97      & 
                                  \cite{tkomatsubara-KTeVpi0mm97}\\
 \hline
 {\ensuremath{\Kp \to \pip \nunub}\xspace} &
        3          &               &
          $1.47^{+1.30}_{-0.89}$ $\times 10^{-10}$ & 
            J  & 
              E949-'02 +E787  & 
                                  \cite{tkomatsubara-E949}\\

 \ \ \ \ \ \ {\ensuremath{\pip \nunub}\xspace}(2) &
        1         & 1.22$\pm$0.24   &
          $<22\times 10^{-10}$  &
            FC& 
              E787      & 
                                  \cite{tkomatsubara-E787pnn2}\\
 \hline
 {\ensuremath{\KL \to \piz \nunub}\xspace} &
         0        & 0.12$^{+0.05}_{-0.04}$ &
          $<5.9\times 10^{-7}$  &
            Poisson   &
              KTeV    & 
                                  \cite{tkomatsubara-KTeVpi0nn}\\
  \cline{2-7}
                                           &
            \multicolumn{4}{c|}{(being analyzed)} &
              E391a-'04        & 
                                   \\
 \hline
  \end{tabular}
\caption{Summary of the studies of FCNC processes.
         This table includes the number of observed events ($N_{obs}$), 
         background estimation ($n_{bgd}$), and the statistical technique used to 
         obtain the result ($Stat$).
         $n_{bgd}$ for {\ensuremath{\Kp \to \pip \nunub}\xspace} is explained in the text.
         ``{\ensuremath{\pip \nunub}\xspace}(2)'' is a search 
          in the \pip\  momentum region $<195$MeV/$c$.
}
\label{tab:tkomatsubara-FCNC}
\end{center}
\end{table}
%%%%%%%%%%%%%%%%%%%%%%%%%%%%%%%%%%%%%%%%%%%%%%%%%%%%%%%%%%%%%%%%%%%%%%%%%

\begin{figure}[htb]
\begin{center}
\epsfig{file=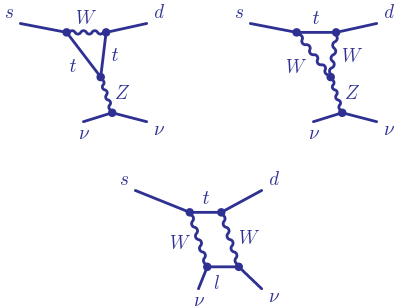,height=4.0cm}
\caption{Penguin and Box diagrams in $K\to\pi\nu\bar{\nu}$.}
\label{fig:tkomatsubara-fig1}
\end{center}
\end{figure}

The FCNC process in kaon decays is
strange-quark to down-quark transition and
is induced in the SM by the $W$ and $Z$ loop effects 
as Penguin and Box diagrams (figure~\ref{fig:tkomatsubara-fig1}).
The top-quark in the loops dominates the transition 
because of its heavy mass, and the quantity: 
\begin{equation}
   V_{ts}^{*} \cdot V_{td} =  
      - A^{2} \lambda^5 \cdot (1 - \rho - i \eta) =
      - |V_{cb}|^{2} \cdot \lambda \cdot (1 - \rho - i \eta)
\label{eq:tkomatsubara-lambdat}
\end{equation}
is measured.  
The decays are rare~\footnote{
  A new $\lambda$~\cite{tkomatsubara-Glazov} that is higher than
  previously thought does not change 
  $V_{ts}^{*} \cdot V_{td}$ so much 
  if $|V_{cb}|$, a directly-measured quantity in B decay, is used.}
due to $|V_{cb}|^{2} \cdot \lambda$, and
are precious because the important parameters $\rho$ and $\eta$
can be determined from them. 
The decay amplitude of \KL\ is 
a superposition of the amplitudes of \Kz\ and \Kzb 
and is proportional to $\eta$; 
observation of rare \KL\ decays, in particular 
the  {\ensuremath{\KL \to \piz \nunub}\xspace} decay, 
is a new evidence for direct CP violation.

%%%%%%%%%%%%%%%%%%%%%%%%%%%%%%%%%%%%%%%%%%%%%%%%%%%%%%%%%%%%%%%%%%%%%%%%%%%
%%
%%   use this format to include an .eps figure into your paper
%%
%%%%%%%%%%%%%%%%%%%%%%%%%%%%%%%%%%%%%%%%%%%%%%%%%%%%%%%%%%%%%%%%%%%%%%%%%%%

 \subsection{{\ensuremath{\Kz \to \piz \ellell}\xspace}}

Rare kaon decays with charged leptons 
should be easier to detect in experiments 
because the \Kz\ invariant mass can be fully-reconstructed.
However, if the kaon decay accompanies 
charged leptons in the final state, 
the transition is also induced 
by long-distance effects with \g\ emission
in hadronic interactions; 
their theoretical interpretations are not straightforward~\footnote{
 The case of the decay {\ensuremath{\KL \to \mup \mun}\xspace}
 (\BR=$(7.27\pm 0.14)\times 10^{-9}$~\cite{tkomatsubara-PDG2004}) 
 is discussed in \cite{tkomatsubara-mmIsidori}.}.
 
To the {\ensuremath{\KL \to \piz \ellell}\xspace} decay, 
there are four contributions
from direct CP violation (DIR),
indirect CP violation 
due to the $K_{1}$ component of \KL\ (MIX),
their interference (INT), 
and the CP conserving process (CPC)
through the \piz{\ensuremath{\gamma^*\gamma^*}\xspace}\ 
intermediate state. 
The CPC contribution can be obtained from  the study of the
decay {\ensuremath{\KL \to \piz \g \g }\xspace}
\cite{tkomatsubara-KLggKTeV,tkomatsubara-KLggNA48}.
The {\ensuremath{\KS \to \piz \ellell}\xspace} decay, which is a CP conserving process of \KS\ , helps
to do reliable estimation of MIX (and INT)
and extract short-distance physics from 
{\ensuremath{\KL \to \piz \ellell}\xspace}
\cite{tkomatsubara-peeBuchalla,tkomatsubara-pmmIsidori}.

\begin{figure}[htb]
\begin{center}
\epsfig{file=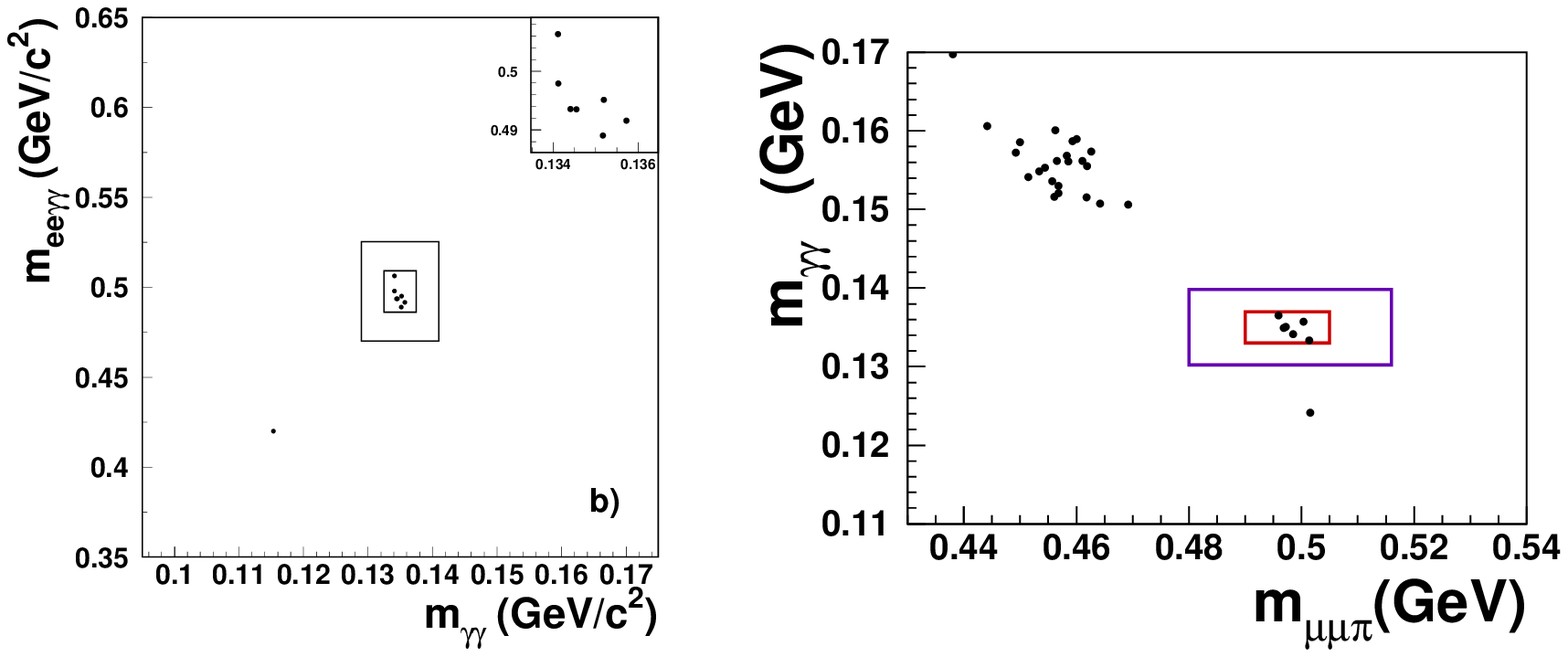,height=5.0cm}
\caption{First observation of
   {\ensuremath{\KS \to \piz \ep \en}\xspace}
   \cite{tkomatsubara-NA48pi0ee} (left) and  
   {\ensuremath{\KS \to \piz \mup \mun}\xspace}
   \cite{tkomatsubara-NA48pi0mm} (right).}
\label{fig:tkomatsubara-fig2}
\end{center}
\end{figure}

 The NA48 collaboration performed the data taking dedicated to 
 \KS decays, named NA48/1, during 89 days in 2002 
 with a high-intensity \KS beam: 
  $2\times 10^5$ \KS decays per spill with a mean energy of 120 GeV.
 The rare decays
 {\ensuremath{\KS \to \piz \ep \en}\xspace} and  
 {\ensuremath{\KS \to \piz \mup \mun}\xspace} 
 were observed for the first time
 (figure~\ref{fig:tkomatsubara-fig2}).
 The background levels 
 (0.15$^{+0.10}_{-0.04}$ and 0.22$^{+0.18}_{-0.11}$, respectively)
 were negligible.
 Using a vector matrix element and unit form factor,
 the measured branching ratios were 
 $5.8^{+2.8}_{-2.3}${\footnotesize (stat)}$\pm 0.8${\footnotesize (syst)} $\times 10^{-9}$ and 
 $2.9^{+1.5}_{-1.2}${\footnotesize (stat)}$\pm 0.2${\footnotesize (syst)} $\times 10^{-9}$, respectively.
 With these results, 
 \BR({\ensuremath{\KL \to \piz \ep \en}\xspace})$\times 10^{12}$ $\simeq$ 
 $5_{DIR}\pm 9_{INT} +17_{MIX}+(negligible)_{CPC}$
 and 
 \BR({\ensuremath{\KL \to \piz \mup \mun}\xspace})$\times 10^{12}$ $\simeq$ 
 $2_{DIR}\pm 3_{INT} +9_{MIX}+5_{CPC}$
 are predicted in the SM; the contribution from DIR 
 is sub-dominant in these decays. 

%%%%%%%%%%%%%%%%%%%%%%%%%%%%%%%%%%%%%%%%%%%%%%%%%%%%%%%%%%%%%%%%%%%%%%%%%%%
%%
%%   use this format to include an .eps figure into your paper
%%
%%%%%%%%%%%%%%%%%%%%%%%%%%%%%%%%%%%%%%%%%%%%%%%%%%%%%%%%%%%%%%%%%%%%%%%%%%%

 The {\ensuremath{\KL \to \piz \ep \en}\xspace} decay
has been studied by KTeV.
The limiting background was 
from the radiative Dalitz decay 
{\ensuremath{\KL \to \ep \en \g \g}\xspace}
(\BR=$(5.95\pm 0.33)\times 10^{-7}$~\cite{tkomatsubara-PDG2004}) 
with invariant mass of the two photons 
consistent with the $\pi^0$ mass~\cite{tkomatsubara-Greenlee}.
Phase space cuts, which were applied to the data to suppress the background, 
reduced the signal acceptance by 25\%. 
The number of events observed in the signal region 
was consistent with the expected background 
for both of their 1997 and 1999 data sets. 
Combining these results, 
the KTeV final result was 
 \BR({\ensuremath{\KL \to \piz \ep \en}\xspace})
$<2.8\times 10^{-10}$.
 KTeV also studied 
 the {\ensuremath{\KL \to \piz \mup \mun}\xspace} decay, 
 and has reported an upper limit 
 \BR({\ensuremath{\KL \to \piz \mup \mun}\xspace})
 $<3.8\times 10^{-10}$ from the 1997 data set
 (and is analyzing the 1999 data set). 
 Both of the limits on \BR({\ensuremath{\KL \to \piz \ep \en}\xspace})
 and  \BR({\ensuremath{\KL \to \piz \mup \mun}\xspace})
 are still an order of magnitude larger than the SM predictions.

 \subsection{{\ensuremath{\kaon \to \pi \nunub}\xspace}}

Branching ratios for the {\ensuremath{\Kp \to \pip \nunub}\xspace}
and {\ensuremath{\KL \to \piz \nunub}\xspace} decays~\cite{tkomatsubara-Buras2004a}
are represented in the SM as
\begin{eqnarray}
 \BR({\ensuremath{\Kp \to \pip \nunub}\xspace}) & = & 
  (\ 5.30 \times 10^{-11}\ )\  \cdot\  C_{\pi \nunub}\ \ \   
    \times\ \ \ [\ (\rho_{0}- \rho )^{2}\ +\ \eta^{2}\ ]
\label{eq:tkomatsubara-kplus}
\end{eqnarray}
and
\begin{eqnarray}
 \BR({\ensuremath{\KL \to \piz \nunub}\xspace}) & = & 
  (\ 23.2 \times 10^{-11}\ )\  \cdot\  C_{\pi \nunub}\ \ \ 
    \times\ \ \ [\ \eta^{2}\ ]\ ,
\label{eq:tkomatsubara-klong}
\end{eqnarray}
respectively, where
\begin{eqnarray}
 C_{\pi \nunub} & \equiv & 
            [\ \frac{\BR({\ensuremath{\Kp \to \piz \ep \nu}\xspace})}{4.87\times 10^{-2}}\ ]\  
    \times\ [\ \frac{|V_{cb}|}{0.0415}\ ]^4\ 
    \times\ [\ \frac{X(x_{t})}{1.529}\ ]^{2}\ ,
\label{eq:tkomatsubara-norm}
\end{eqnarray}
$X(x_{t})$ is the Inami-Lim loop function~\cite{tkomatsubara-InamiLim}
with the QCD correction,
$x_t$ is the square of the ratio of the top to W masses.
$\rho_0$ in \BR({\ensuremath{\Kp \to \pip \nunub}\xspace})~\footnote{
 The departure of $\rho_0$ from unity,  
 without which the branching ratio
 should be proportional to $|V_{td}|^2$, 
 measures the relative importance of the internal charm-quark contributions.} 
is estimated to be $\approx 1.37$.
Long-distance contributions are negligible, and the hadronic matrix elements 
are extracted from the 
{\ensuremath{\Kp \to \piz \ep \nu}\xspace} decay.
The theoretical uncertainty in \BR({\ensuremath{\Kp \to \pip \nunub}\xspace})
is 7\%
from the charm-quark contribution in the next-to-leading logarithmic(NLO) QCD calculations
in $\rho_0$ and would be reduced by performing a next-to-NLO calculation, while 
the theoretical uncertainty in  \BR({\ensuremath{\KL \to \piz \nunub}\xspace})
is only 1-2\%.
With the $\rho$-$\eta$ constraints from other kaon and B decay experiments, 
the SM prediction is 
$(7.8\pm 1.2)\times 10^{-11}$ for \BR({\ensuremath{\Kp \to \pip \nunub}\xspace})
and $(3.0\pm 0.6)\times 10^{-11}$ for \BR({\ensuremath{\KL \to \piz \nunub}\xspace}).
New physics beyond the SM could affect 
these branching ratios~\cite{tkomatsubara-Buras2001,tkomatsubara-Buras2004b,tkomatsubara-Buras2004c},
and the $\rho$ and $\eta$ (and $\sin 2\phi_1$) determined from  
{\ensuremath{\kaon \to \pi \nunub}\xspace} 
and those from the B system would be different~\cite{tkomatsubara-Belyaev}.
Since the effects of new physics are not expected to be too large,
a precise measurement of a decay at the level of $10^{-11}$ is required.

The E787 and E949 collaborations for
{\ensuremath{\Kp \to \pip \nunub}\xspace} and related decays 
measure the charged track emanating from $K^+$ decays at rest.
The $\pi^+$ momentum from $K^+\to\pi^+\nu\bar{\nu}$
is less than 227MeV/$c$, 
while the major background sources of
$K^+\to\pi^+\pi^0$ ($K^+_{\pi 2}$, \BR=21.2\%) and
$K^+\to\mu^+\nu$   (\BR=63.5\%)
are two-body decays and have monochromatic momentum of
205MeV/$c$ and 236MeV/$c$, respectively.
The region ``above the $K^+_{\pi 2}$'' between 211MeV/$c$ and 229MeV/$c$
is adopted for the search~\footnote{
   E787 reported a new upper limit
   \BR({\ensuremath{\Kp \to \pip \nunub}\xspace})
   $<22\times 10^{-10}$ 
   from a search for {\ensuremath{\Kp \to \pip \nunub}\xspace}
          in the \pip\  momentum region $<195$MeV/$c$.
   E949 continues the search in the same region.}.
Background rejection is essential in this experiment,
and the weapons for
redundant kinematics measurement,
$\mu^+$ rejection, and extra-particle and photon veto
are employed.

%%%%%%%%%%%%%%%%%%%%%%%%%%%%%%%%%%%%%%%%%%%%%%%%%%%%%%%%%%%%%%%%%%%%%%%%%%%
%%
%%   use this format to include an .eps figure into your paper
%%
\begin{figure}[htb]
\begin{center}
\epsfig{file=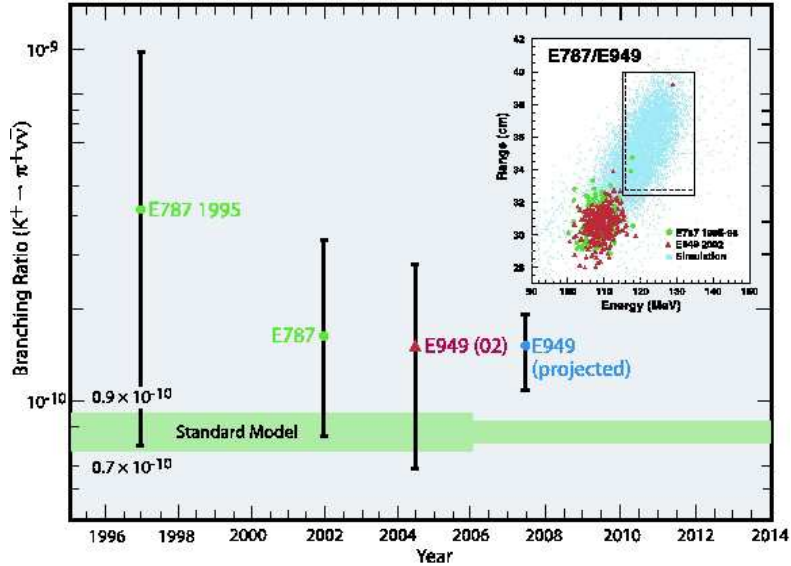,height=7.5cm}
\caption{History of the study of {\ensuremath{\Kp \to \pip \nunub}\xspace},
   with the plot~\cite{tkomatsubara-E949} of E949-'02 +E787 at the top right.
 ``E949 (02)'' represents the branching ratio by combining the E949 and E787 data sets; 
 ``E949 (projected)'' represents the measured central value
 with the precision expected to E949 after running for 60 weeks.
 The SM prediction would be narrowed down
 if the \Bs-\Bsb mixing is measured in the near future.}
\label{fig:tkomatsubara-fig3}
\end{center}
\end{figure}
%%%%%%%%%%%%%%%%%%%%%%%%%%%%%%%%%%%%%%%%%%%%%%%%%%%%%%%%%%%%%%%%%%%%%%%%%%%

 E787 had reported two events consistent with 
 the {\ensuremath{\Kp \to \pip \nunub}\xspace} decay 
 giving \BR({\ensuremath{\Kp \to \pip \nunub}\xspace})
 $=1.57^{+1.75}_{-0.82}$ $\times 10^{-10}$ 
 with the 1995-1998 data sets~\cite{tkomatsubara-E787}.
 The backgrounds were estimated to contribute $0.15\pm 0.05$ events.
 In the first data set of E949 for 12 weeks in 2002
 with the upgraded detector~\cite{tkomatsubara-E949Web}, 
 an additional event near the upper kinematic limit for 
 {\ensuremath{\Kp \to \pip \nunub}\xspace} was observed. 
 A total number of background events expected in the signal region
 was $0.30\pm 0.03$.
 Combining the E949 and E787 data sets, the branching ratio
 \BR({\ensuremath{\Kp \to \pip \nunub}\xspace})
 $=1.47^{+1.30}_{-0.89}$ $\times 10^{-10}$,
 in the 68\% confidence interval including statistical and systematic 
 uncertainties, 
 was obtained by a likelihood ratio technique 
 based on the three observed events (figure~\ref{fig:tkomatsubara-fig3}). 
 At the measured central value of the branching ratio, 
 the additional event had a signal-to-background ratio of 0.9; 
 the estimated probability that background alone gave rise to the three events
 (or to any more signal-like configuration) was 0.001. 
 The upper limit for \BR({\ensuremath{\Kp \to \pip \nunub}\xspace}) was
 $<3.22\times 10^{-10}$, from which 
 a model-independent bound on {\ensuremath{\KL \to \piz \nunub}\xspace}: 
 $\BR({\ensuremath{\KL \to \piz \nunub}\xspace}) < 
    4.4 \times \BR({\ensuremath{\Kp \to \pip \nunub}\xspace})
   < 1.4 \times 10^{-9}$
 (Grossman-Nir limit~\cite{tkomatsubara-GN}) can be extracted. 
 The E949 and E787 data sets were also used to set a limit on 
 the branching ratio for {\ensuremath{\Kp \to \pip X^0}\xspace}, 
 where $X^0$ is a neutral weakly-interacting massless particle~\cite{tkomatsubara-familon},
 to be  $< 7.3 \times 10^{-11}$
 based on the one event from E949~\footnote{This event was observed within 2 standard deviations
 of the expected \pip\ momentum (227MeV/$c$).}.
 E949 was approved to run for 60 weeks
 and is waiting to take more data. A cosmic-ray run with the E949 detector
 was performed in August 2004,  and the collaboration is ready to resume
 {\ensuremath{\pip \nunub}\xspace} data collection.

The current best upper limit on 
\BR({\ensuremath{\KL \to \piz \nunub}\xspace})
$< 5.9 \times 10^{-7}$ was set 
by KTeV; 
the Dalitz decay mode $\piz\to e^+e^-\gamma$ (\BR=1.2\%)
for the final state of {\ensuremath{\KL \to \piz \nunub}\xspace}
was used in this search.
To reach the SM sensitivity, 
reconstructing $\piz\to \gamma\gamma$ from {\ensuremath{\KL \to \piz \nunub}\xspace}
has to be considered.
To beat the major background from $K^0_L\to\pi^0\pi^0$
in case two out of four photons are missed, 
photon detection with low inefficiency ($< 10^{-3}\sim 10^{-4}$) is
required to the detector. 
A search using $\pi^0\to \gamma\gamma$
had been performed
with KTeV's one-day special run~\cite{tkomatsubara-KTeV1day}; 
a limit on the branching ratio was determined to be 
$< 1.6 \times 10^{-6}$
based on one event in the signal region 
with the background estimate of $3.5\pm 0.9$ events.

%%%%%%%%%%%%%%%%%%%%%%%%%%%%%%%%%%%%%%%%%%%%%%%%%%%%%%%%%%%%%%%%%%%%%%%%%%%
%%
%%   use this format to include an .eps figure into your paper
%%
\begin{figure}[htb]
\begin{center}
\epsfig{file=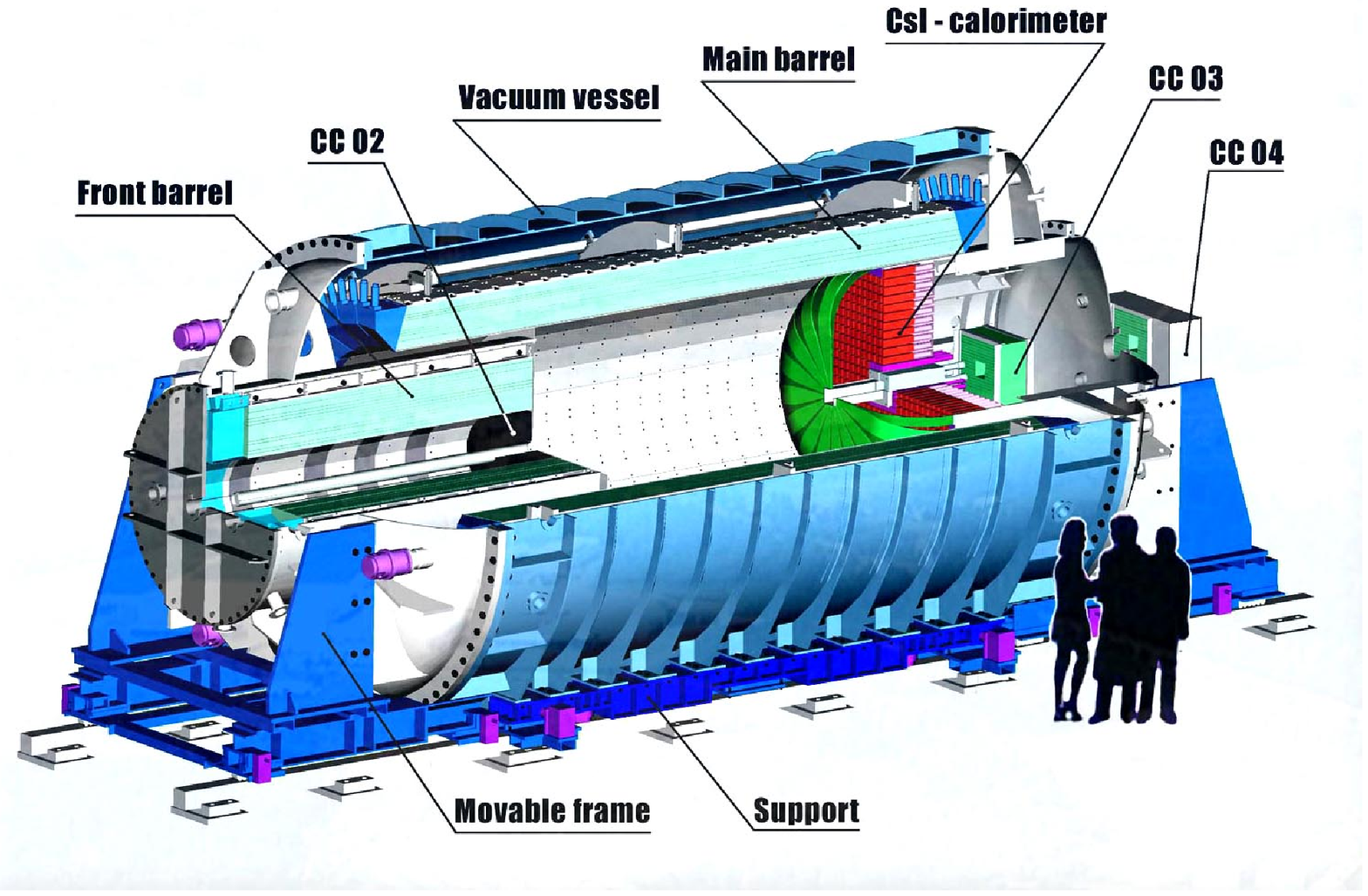,height=7.5cm}
\caption{E391a detector for {\ensuremath{\KL \to \piz \nunub}\xspace}.}
\label{fig:tkomatsubara-fig4}
\end{center}
\end{figure}
%%%%%%%%%%%%%%%%%%%%%%%%%%%%%%%%%%%%%%%%%%%%%%%%%%%%%%%%%%%%%%%%%%%%%%%%%%%

The E391a experiment~\cite{tkomatsubara-E391aWeb} (figure~\ref{fig:tkomatsubara-fig4})
is the first dedicated search
for the {\ensuremath{\KL \to \piz \nunub}\xspace} decay.
A collimated ``pencil'' neutral beam was newly designed, and
an endcap calorimeter with undoped CsI crystals 
detected two photons from \piz
and measured their energy and position.
The \KL-decay vertex position along the beam line 
was determined from the constraint of \piz mass. 
Calorimeters that covered the decay region did 
hermetic photon detection.
Charged particles were removed by their energy deposits in plastic scintillators 
in front of each calorimeter. 
Beam line survey and detector construction were performed 
from 2001 to 2003 and the first physics run was carried out in 2004
with $2.5\times 10^{12}$ protons per spill. 
The goal of E391a is to achieve a sensitivity 
below the Grossman-Nir limit 
and to reach the level predicted by 
new physics~\cite{tkomatsubara-Buras2004b,tkomatsubara-Buras2004c}; 
the 2004 data set is now being analyzed, and the next physics run is scheduled in 2005.

\section{Future kaon programs}

 All the kaon laboratories have their own project of 
 high-intensity proton accelerator facilities: 
 J-PARC of KEK-JAERI, AGS of BNL, SPS of CERN and Main Injector
 of FNAL. 
 The workshops, in which 
 new kaon programs were discussed earnestly, were held in the last 6 months; 
 please visit the Web sites~\cite{tkomatsubara-NP04,tkomatsubara-BNL,tkomatsubara-HIF04,tkomatsubara-PDW}
 for details. 

 J-PARC, which stands for Japan Proton Accelerator Research Complex~\cite{tkomatsubara-JPARC},
 in the joint project of Japan Atomic Energy Research Institute (JAERI) and KEK. 
 The accelerators consisting of Linac, 3GeV rapid-cycle Synchrotron and 
 50GeV Synchrotron are under construction at the Tokai site of JAERI
 located at 50km northeast of KEK. The construction of the Phase-1 facilities 
 will be finished in 2008 and, with the very intense proton beam
 ($300\times 10^{12}$ protons per spill)
 from the 50GeV PS, great opportunities for various research
 in nuclear and particle physics, including kaon experiments
 with much higher sensitivities than ever, would be opened. 
 A call for Letters of Intent (LoI's) was issued in July 2002,
 and thirty LoI's were submitted.
 There were five LoI's for kaon experiments 
 with a neutral beam and with a \Kp\ beam of low momentum (0.6-0.8 GeV/$c$); 
 these are regarded as a natural extension of the kaon experiments
 that have been worked out (E391a, E949 and E246).
 See \cite{tkomatsubara-DAFNE2004}
 for more information.

  KOPIO~\cite{tkomatsubara-KOPIO},
 which is a part of the Rare Symmetry Violating Processes (RSVP)
 experiments,
 is a study of {\ensuremath{\KL \to \piz \nunub}\xspace}
 with new concepts and techniques. 
 An RF-bunched proton beam from BNL-AGS with a large targeting angle
 produces \KL's of low momentum (around 0.8 GeV/$c$), 
 so that with the TOF technique 
 the momentum of each kaon is measured 
 and the decay can be analyzed 
 in the \KL\ rest frame.
 A combination of the pre-radiator and Shashlik calorimeter
 intends to measure the timing, energy, position and angle
 of low energy photons and fully reconstruct the decay. 
 The RSVP construction start is in the FY05 Congressional budget 
 of the US, and the first physics run is expected to be performed 
 in 2010. 

 There is a FNAL experiment named CKM~\cite{tkomatsubara-CKM}
 to construct an RF-separated \Kp beam of 22 GeV/$c$ from Main Injector 
 and measure the \pip \nunub\ decay in flight for the first time. 
 The CKM experiment was granted scientific (Stage 1) approval in 2001
 but was not endorsed by the P5 Subpanel of HEPAP for cost reasons; 
 the experiment is being re-designed as P940 to use 
 an un-separated  \Kp beam of 45 GeV/$c$ to the KTeV experimental hall. 
 At CERN-SPS a new initiative called NA48/3~\cite{tkomatsubara-NA48-3},
 which uses an un-separated \Kp beam of 75 GeV/$c$ 
 for the {\ensuremath{\Kp \to \pip \nunub}\xspace} decay in flight,
 is in progress.
 Beam tests for the detectors in NA48/3
 were performed in August 2004, and 
 data to address the issues of drift chambers, 
 a new beam spectrometer (``gigatracker'')
 based on hybrid technology with thin-and-fast silicon
 micro-pixel layers plus a micromegas-based TPC,
 and photon hermeticity
 were collected. Their LoI (SPSC-I-229) was submitted in October.

\section{Conclusions}

 The study of kaon physics continues to make great strides. 
 Though no explicit violation of the SM is observed,
 certain new-physics scenarios have been excluded.
 Experimental sensitivities have reached $10^{-9}$-$10^{-12}$.
 ``Blind analysis''
 has already been de facto; 
 in the rare-decay searches where both Signal and Noise are in small statistic, 
 we started using Likelihood techniques 
 instead of a simple cut-and-count method. 
 Future kaon programs will be ``almost {\ensuremath{\pi \nunub}\xspace}''~\footnote{
  Right now there is no initiative for new LF-violation experiments in kaon decays.}; 
 \BR({\ensuremath{\Kp \to \pip \nunub}\xspace})
and
 \BR({\ensuremath{\KL \to \piz \nunub}\xspace})
would be measured  
in $\sim$10\% precision ($\sim$100 signal events)
with highly sophisticated and special-purpose detectors,
and it will be tested 
whether the source of CP violation is only from the CKM phase or not.

\section*{Acknowledgments}
I would like to thank 
  J.A. Appel,
  L. Bellantoni, A. Bellavance,
  F. Bossi,  
  D. A. Bryman, 
  A. Ceccucci, 
  S. Glazov,
  X-G. He, 
  J. Imazato, 
  T. Inagaki, 
  S.H. Kettell, 
  G.Y. Lim, 
  L.S. Littenberg, 
  I. Mikulec,
  G. Redlinger, 
  A. Sher, 
  S. Sugimoto, 
and
  R. Tschirhart
for providing me help with my talk
at the FPCP2004 conference and this article for the Proceedings. 
I would like to acknowledge support from Grant-in-Aid for Scientific Research
in Priority Area: ``Mass Origin and Supersymmetry Physics'' 
by the MEXT Ministry in Japan.

%%%%%%%%%%%%%%%%%%%%%%%%%%%%%%%%%%%%%
% Label to flag the last page of your contribution
% Replace Perret by your name starting with a capital letter
%
\label{TKomatsubaraEnd}

\end{document}

%% file: fpcp04-macro.tex
\def\beq{\begin{equation}}
\def\eeq#1{\label{#1}\end{equation}}
\def\eeqn{\end{equation}}

\def\beqa{\begin{eqnarray}}
\def\eeqa#1{\label{#1}\end{eqnarray}}
\def\eeqan{\end{eqnarray}}

\let\bar=\overbar

\def\Dslash{\not{\hbox{\kern-4pt $D$}}}
\def\dslash{\not{\hbox{\kern-2pt $\del$}}}

\def\msb{{\bar{\ssstyle M \kern -1pt S}}}

\def\BB0bar{B^0 {\overline B}^0}
\def\BB0dbar{B_d^0 {\overline B}_d^0}
\def\BB0sbar{B_s^0 {\overline B}_s^0}

\RequirePackage{xspace}

\usepackage{relsize}
\def\babar{\mbox{\slshape B\kern-0.1em{\smaller A}\kern-0.1em
    B\kern-0.1em{\smaller A\kern-0.2em R}}}

\def\en         {\ensuremath{e^-}\xspace}   % electron negative (\em is taken)
\def\ep         {\ensuremath{e^+}\xspace}
 
\def\epem       {\ensuremath{e^+e^-}\xspace}

\def\mup        {\ensuremath{\mu^+}\xspace}
\def\mun        {\ensuremath{\mu^-}\xspace} % muon negative (\mum is taken)

\def\ellell     {\ensuremath{\ell^+ \ell^-}\xspace}

\def\nunub      {\ensuremath{\nu{\overline{\nu}}}\xspace}

\def\nunub      {\ensuremath{\nu{\overline{\nu}}}\xspace}

\def\g     {\ensuremath{\gamma}\xspace}
\def\piz   {\ensuremath{\pi^0}\xspace}

\def\pip   {\ensuremath{\pi^+}\xspace}
\def\pim   {\ensuremath{\pi^-}\xspace}

\def\kaon  {\ensuremath{K}\xspace}
\def\Kbar  {\kern 0.2em\overline{\kern -0.2em K}{}\xspace}

\def\Kz    {\ensuremath{K^0}\xspace}
\def\Kzb   {\ensuremath{\Kbar^0}\xspace}
\def\KzKzb {\ensuremath{\Kz \kern -0.16em \Kzb}\xspace}
\def\Kp    {\ensuremath{K^+}\xspace}
\def\Km    {\ensuremath{K^-}\xspace}

\def\KpKm  {\ensuremath{\Kp \kern -0.16em \Km}\xspace}
\def\KS    {\ensuremath{K^0_{\scriptscriptstyle S}}\xspace} 
\def\KL    {\ensuremath{K^0_{\scriptscriptstyle L}}\xspace}

\def\Dbar    {\kern 0.2em\overline{\kern -0.2em D}{}\xspace}

\def\Dz      {\ensuremath{D^0}\xspace}
\def\Dzb     {\ensuremath{\Dbar^0}\xspace}
\def\DzDzb   {\ensuremath{\Dz {\kern -0.16em \Dzb}}\xspace}
\def\Dp      {\ensuremath{D^+}\xspace}
\def\Dm      {\ensuremath{D^-}\xspace}

\def\DpDm    {\ensuremath{\Dp {\kern -0.16em \Dm}}\xspace}

\def\Bbar    {\kern 0.18em\overline{\kern -0.18em B}{}\xspace}

\def\BB      {\ensuremath{B\Bbar}\xspace} 
\def\Bz      {\ensuremath{B^0}\xspace}
\def\Bzb     {\ensuremath{\Bbar^0}\xspace}
\def\BzBzb   {\ensuremath{\Bz {\kern -0.16em \Bzb}}\xspace}
\def\Bu      {\ensuremath{B^+}\xspace}
\def\Bub     {\ensuremath{B^-}\xspace}

\def\BpBm    {\ensuremath{\Bu {\kern -0.16em \Bub}}\xspace}
\def\Bs      {\ensuremath{B_s}\xspace}
\def\Bsb     {\ensuremath{\Bbar_s}\xspace}

\mathchardef\Upsilon="7107
\def\Y#1S{\ensuremath{\Upsilon{(#1S)}}\xspace}% no space before {...}!

\mathchardef\Deltares="7101
\mathchardef\Xi="7104
\mathchardef\Lambda="7103
\mathchardef\Sigma="7106
\mathchardef\Omega="710A

\def\Deltabar{\kern 0.25em\overline{\kern -0.25em \Deltares}{}\xspace}
\def\Lbar{\kern 0.2em\overline{\kern -0.2em\Lambda\kern 0.05em}\kern-0.05em{}\xspace}
\def\Sigbar{\kern 0.2em\overline{\kern -0.2em \Sigma}{}\xspace}
\def\Xibar{\kern 0.2em\overline{\kern -0.2em \Xi}{}\xspace}
\def\Obar{\kern 0.2em\overline{\kern -0.2em \Omega}{}\xspace}
\def\Nbar{\kern 0.2em\overline{\kern -0.2em N}{}\xspace}
\def\Xb{\kern 0.2em\overline{\kern -0.2em X}{}\xspace}

\def\BR         {{\ensuremath{\cal B}\xspace}}

\newcommand{\tev}{\ensuremath{\mathrm{\,Te\kern -0.1em V}}\xspace}
\newcommand{\gev}{\ensuremath{\mathrm{\,Ge\kern -0.1em V}}\xspace}
\newcommand{\mev}{\ensuremath{\mathrm{\,Me\kern -0.1em V}}\xspace}
\newcommand{\kev}{\ensuremath{\mathrm{\,ke\kern -0.1em V}}\xspace}
\newcommand{\ev}{\ensuremath{\mathrm{\,e\kern -0.1em V}}\xspace}
\newcommand{\gevc}{\ensuremath{{\mathrm{\,Ge\kern -0.1em V\!/}c}}\xspace}
\newcommand{\mevc}{\ensuremath{{\mathrm{\,Me\kern -0.1em V\!/}c}}\xspace}
\newcommand{\gevcc}{\ensuremath{{\mathrm{\,Ge\kern -0.1em V\!/}c^2}}\xspace}
\newcommand{\mevcc}{\ensuremath{{\mathrm{\,Me\kern -0.1em V\!/}c^2}}\xspace}
\def\mus  {\ensuremath{\rm \,\mus}\xspace}

\def\mus        {\ensuremath{\,\mu{\rm s}}\xspace}    %% microsecond
\def\to                 {\ensuremath{\rightarrow}\xspace}

\def\pep2{PEP-II}

\def\gsim{{~\raise.15em\hbox{$>$}\kern-.85em
          \lower.35em\hbox{$\sim$}~}\xspace}
\def\lsim{{~\raise.15em\hbox{$<$}\kern-.85em
          \lower.35em\hbox{$\sim$}~}\xspace}

\xspace

\def\jetset74   {\mbox{\tt Jetset \hspace{-0.5em}7.\hspace{-0.2em}4}\xspace}

%% file: tkomatsubara_arXiv.bbl
\begin{thebibliography}{99}

%%
%%  bibliographic items can be constructed using the LaTeX format in SPIRES:
%%    see    http://www.slac.stanford.edu/spires/hep/latex.html
%%  SPIRES will also supply the CITATION line information; please include it.
%%



\bibitem{tkomatsubara-Bossi}F. Bossi,
 {\it these proceedings}
\bibitem{tkomatsubara-Glazov}S. Glazov,
 {\it these proceedings}
\bibitem{tkomatsubara-He}X-G. He, 
 {\it these proceedings}
\bibitem{tkomatsubara-E871ee}D. Ambrose {\it et al.}, 
   Phys. Rev. Lett. {\bf 81}, 4309 (1998)
\bibitem{tkomatsubara-E871mue}D. Ambrose {\it et al.}, 
   Phys. Rev. Lett. {\bf 81}, 5734 (1998)
\bibitem{tkomatsubara-PDG2004litt}L. Littenberg and G. Valencia in 
    ``Review of Particle Physics'', 
   Phys. Lett. {\bf B592}, 607 (2004)
\bibitem{tkomatsubara-Redlinger}G. Redlinger, presented at DA$\Phi$NE2004, hep-ex/0408142 
\bibitem{tkomatsubara-Mikulec}I. Mikulec, presented at XXIV Physics in Collision, hep-ex/0409050
\bibitem{tkomatsubara-Patera}V. Patera, presented at ICHEP'04

\bibitem{tkomatsubara-FC}G.J. Feldman and R.D. Cousins, 
   Phys. Rev. {\bf D57}, 3873 (1998)

\bibitem{tkomatsubara-Sher}A. Sher, presented at DPF2004
\bibitem{tkomatsubara-Bellavance}A. Bellavance, private communication
\bibitem{tkomatsubara-E246}M. Abe {\it et al.}, 
   Phys. Rev. Lett. {\bf 93}, 131601 (2004)
\bibitem{tkomatsubara-E865ex}R. Appel {\it et al.}, 
   Phys. Rev. Lett. {\bf 85}, 2877 (2000)
\bibitem{tkomatsubara-E787pig}S. Adler {\it et al.}, 
   Phys. Rev. {\bf D65}, 052009 (2002)
\bibitem{tkomatsubara-E949}V.V. Anisimovsky {\it et al.}, 
   Phys. Rev. Lett. {\bf 93}, 031801 (2004)
\bibitem{tkomatsubara-LFVtheory1}R.N. Cahn and H. Harari, 
   Nucl. Phys. {\bf B176}, 135 (1980)
\bibitem{tkomatsubara-LFVtheory2} O. Shanker, 
   Nucl. Phys. {\bf B206}, 253 (1982)
\bibitem{tkomatsubara-SUSY} B.A. Campbell, 
   Phys. Rev. {\bf D28}, 209 (1983)
\bibitem{tkomatsubara-LFVtheory3} A. Belyaev {\it et al.},
   Eur.Phys.J. {\bf C22}, 715 (2002)
\bibitem{tkomatsubara-LFVtheory4} J.-M. Fr\'{e}re {\it et al.},
   JHEP {\bf 0403}, 001 (2004)
\bibitem{tkomatsubara-LFVtheory5} T. Appelquist {\it et al.},
   Phys. Rev. {\bf D69}, 015002 (2004)
\bibitem{tkomatsubara-LFVtheory6} L.G. Landsberg,
   hep-ex/0410261
\bibitem{tkomatsubara-E777}A.M. Lee {\it et al.}, 
   Phys. Rev. Lett. {\bf 64}, 165 (1990)
\bibitem{tkomatsubara-PDG2004}Particle Data Group: S. Eidelman {\it et al.}, 
    ``Review of Particle Physics'', 
   Phys. Lett. {\bf B592}, 1 (2004)
\bibitem{tkomatsubara-Sanda}I.I. Bigi and A.I. Sanda, 
  ``CP violation'' (Cambridge University Press, 2000)
\bibitem{tkomatsubara-LittShrock1} L.S. Littenberg and R.E. Shrock, 
   Phys. Rev. Lett. {\bf 68}, 443 (1992)
\bibitem{tkomatsubara-Zuber} K. Zuber,
   Phys. Lett. {\bf B479}, 33 (2000)
\bibitem{tkomatsubara-LittShrock2} L.S. Littenberg and R.E. Shrock, 
   Phys. Lett. {\bf B491}, 285 (2000)
\bibitem{tkomatsubara-exotic}J. Trampeti\'{c}, hep-ph/0212309

\bibitem{tkomatsubara-Junk}T. Junk, 
   Nucl. Instrum. Methods Phys. Res., Sect. A {\bf 434}, 435 (1999)


\bibitem{tkomatsubara-NA48pi0ee}J.R. Batley {\it et al.}, 
   Phys. Lett. {\bf B576}, 43 (2003)
\bibitem{tkomatsubara-NA48pi0mm}J.R. Batley {\it et al.}, 
   Phys. Lett. {\bf B599}, 197 (2004)
\bibitem{tkomatsubara-KTeVpi0ee97}A. Alavi-Harati {\it et al.}, 
   Phys. Rev. Lett. {\bf 86}, 397 (2001)
\bibitem{tkomatsubara-KTeVpi0ee99}A. Alavi-Harati {\it et al.}, 
   Phys. Rev. Lett. {\bf 93}, 021805 (2004)
\bibitem{tkomatsubara-KTeVpi0mm97}A. Alavi-Harati {\it et al.}, 
   Phys. Rev. Lett. {\bf 84}, 5279 (2000)

\bibitem{tkomatsubara-E787pnn2}S. Adler {\it et al.}, 
   Phys. Rev. {\bf D70}, 037102 (2004)
\bibitem{tkomatsubara-KTeVpi0nn}A. Alavi-Harati {\it et al.}, 
   Phys. Rev. {\bf D61}, 072006 (2000)
\bibitem{tkomatsubara-mmIsidori} G. Isidori and R. Unterdorfer, 
   JHEP {\bf 0401}, 009 (2004)
\bibitem{tkomatsubara-KLggKTeV}A. Alavi-Harati {\it et al.}, 
   Phys. Rev. Lett. {\bf 83}, 917 (1999)
\bibitem{tkomatsubara-KLggNA48}A. Lai {\it et al.}, 
   Phys. Lett. {\bf B536}, 229 (2002)
\bibitem{tkomatsubara-peeBuchalla} G. Buchalla {\it et al.},
   Phys. Lett. {\bf B672}, 387 (2003)
\bibitem{tkomatsubara-pmmIsidori} G. Isidori {\it et al.},
   Eur.Phys.J. {\bf C36}, 57 (2004)
\bibitem{tkomatsubara-Greenlee}H.B. Greenlee,
   Phys. Rev. {\bf D42}, 3724 (1990)
\bibitem{tkomatsubara-Buras2004a}A.J. Buras {\it et al.},
   hep-ph/0405132, and references therein
\bibitem{tkomatsubara-InamiLim}T. Inami and C.S. Lim, 
   Progr. Theor. Phys. {\bf 65}, 297 (1981); 1172(E) (1981)
\bibitem{tkomatsubara-Buras2001}A.J. Buras and R. Fleischer, 
   Phys. Rev. {\bf D64}, 115010 (2001)
\bibitem{tkomatsubara-Buras2004b}A.J. Buras {\it et al.},
   Phys. Rev. Lett. {\bf 92}, 101804 (2004); 
   Nucl. Phys. {\bf B697}, 133 (2004)
\bibitem{tkomatsubara-Buras2004c}A.J. Buras {\it et al.},
   hep-ph/0408142
\bibitem{tkomatsubara-Belyaev}A. Belyaev {\it et al.},
   hep-ph/0107046
\bibitem{tkomatsubara-E787}S. Adler {\it et al.}, 
   Phys. Rev. Lett. {\bf 88}, 041803 (2002)
\bibitem{tkomatsubara-E949Web}
   {\tt http://www.phy.bnl.gov/e949/}
\bibitem{tkomatsubara-GN}Y. Grossman and Y. Nir, 
   Phys. Lett. {\bf B398}, 163 (1997)
\bibitem{tkomatsubara-familon}F. Wilczek, 
   Phys. Rev. Lett. {\bf 49}, 1549 (1982)

\bibitem{tkomatsubara-KTeV1day}J. Adams {\it et al.},
   Phys. Lett. {\bf B447}, 240 (1999)
\bibitem{tkomatsubara-E391aWeb}
   {\tt http://www-ps.kek.jp/e391/}
\bibitem{tkomatsubara-NP04}
   NP04: The 3rd International Workshop on Nuclear and Particle Physics 
   at J-PARC, {\tt http://j-parc.jp/NP04/}
\bibitem{tkomatsubara-BNL}
   Workshop on Future Kaon Experiments at the AGS, 
   {\tt http://www3.bnl.gov/FutureK/}
\bibitem{tkomatsubara-HIF04}
   HIF04: High Intensity Frontier Workshop, 
   {\tt http://www.pi.infn.it/pm/2004/}
\bibitem{tkomatsubara-PDW}
   Fermilab Proton Driver Workshop, \\
   {\tt http://www-td.fnal.gov/projects/PD/PhysicsIncludes/Workshop/}
\bibitem{tkomatsubara-JPARC}
   {\tt http://j-parc.jp/}
\bibitem{tkomatsubara-DAFNE2004} T.K. Komatsubara, 
   hep-ex/0409017, and references therein
\bibitem{tkomatsubara-KOPIO}
   {\tt http://www.kopio.bnl.gov/}
\bibitem{tkomatsubara-CKM}
   {\tt http://www.fnal.gov/projects/ckm/Welcome.html}
\bibitem{tkomatsubara-NA48-3}
   {\tt http://na48.web.cern.ch/NA48/NA48-3/}


\end{thebibliography}
